# Electromyography biofeedback system with visual and vibratory feedbacks designed for lower limb rehabilitation

João Vitor da Silva Moreira, Karina A. Rodrigues, Daniel José L.L. Pinheiro, Thaís C. Santos, João Luiz Vieira, Esper A. Cavalheiro, Jean Faber

*Laboratory of Neuroengineering and Neurocognition, Department of Neurology and Neurosurgery, Federal University of São Paulo, São Paulo–SP Brazil*

**Abstract**— One of the main causes of long-term prosthetic abandonment is the lack of ownership over the prosthesis, caused mainly by the absence of sensory information regarding the lost limb. One strategy to overcome this problem is to provide alternative feedback mechanisms to convey information respective to the absent limb. To address this issue, we developed a Biofeedback system for the rehabilitation of transfemoral amputees, controlled via electromyographic activity from the leg muscles, that can provide real-time visual and/or vibratory feedback for the user. In this study, we tested this device with able-bodied individuals performing an adapted version of the clinical protocol. Our idea was to test the effectiveness of combining vibratory and visual feedbacks and how task difficulty affects overall performance. Our results show no negative interference combining both feedback modalities, and that performance peaked at the intermediate difficulty. These results provide powerful insights of what can be expected with the population of amputee people and will help in the final steps of protocol development. Our goal is to use this biofeedback system to engage another sensory modality in the process of spatial representation of a virtual leg, bypassing the lack of information associated with the disruption of afferent pathways following amputation.

**Key Terms**— Biofeedback, Electromyography, Virtual Reality, Vibratory Feedbacks, Lower Limb Rehabilitation, Crossmodal feedback.

## 1. Introduction

Limb amputations have a great impact on a person's life, from attitudinal barriers, to physical [1], psychological [1] and even changes in brain functioning [2,3]. Current post-surgical rehabilitation strategies focus mainly on the clinical aspects involved in this process, such as pain control, wound healing, maintenance of muscle strength and joint mobility in the stump [1,4]. One of the most important periods in limb loss rehabilitation is the preprosthetic training, where the patient learns how to interact with a prosthetic device [1,4]. It ideally happens within the first months after amputation, which is also a period associated with the consolidation of brain changes that can further lead to Phantom Limb Phenomena [2–4]. The introduction of feedback mechanisms is critical to bypass the lack of sensorial information between the user and the prosthesis, especially when the idea is to reduce device abandonment [5,6]. Biofeedback systems can provide such bypass and are one of the most promising technologies to promote rehabilitation.

A Biofeedback system refers both to a process, that involves learning associated with feedback stimulus, and a device, which translates physiological information into sensory feedback responses [7]. Feedback information is transmitted either via modality-matched (e.g. a touch on the prosthesis is felt as a touch sensation) or by sensory substitution (or Crossmodal feedback, e.g. joint angle represented as a varying tone) [6, 8–10], and some common feedback types are visual [11], auditory [12], tactile [8,13] and haptic [14].

Learning with Biofeedback occurs upon training and is guided by the model of operant conditioning [15]. It works by selectively reinforcing the responses of interest, while gradually progressing the specificity of this response, achieved by increasing the difficulty of the proposed task [16]. Thus, users learn to modulate a physiological response of interest [17]. Recent studies show its efficacy in a variety of conditions [12,18,19], including lower limb amputations [20]. More importantly, literature suggests that the association of different modalities provides a better mechanism for feedback interpretation and learning [21], given the natural ability of the brain to integrate multisensory information [22].

In this study, we report the development of a Biofeedback that provides both, visual and vibratory feedback to the user, and we evaluated the response of these feedbacks in a group of able-bodied individuals (with no amputation) in a spatial representation task. The goal was to test the ability of the users to assimilate the feedback modalities as source of information. The system uses surface electromyographic (EMG) activity to control the movements of a prosthetic leg in a Virtual Reality environment (VR) and the activation of a matrix of vibrotactile actuators in the back of the user, and the term spatial representation refers to the reproduction of the proposed movements either via the position of the virtual leg or the vibration of the actuators in the back.

The system is built upon the operant conditioning paradigm [16], which is addressed as: (*i*) the influence of contingent feedback in the consolidation of the relationship between vibratory feedback and the spatial localization of a virtual leg; (*ii*) positive and negative reinforcements provided by auditory beeps after each trial during the tasks, and more importantly, (*iii*) a gradual increase in task difficulty, (*iv*) and the balance between the levels of activation of the agonist and antagonist muscles (an aspect that was not explored in this particular study) [16].

During the tests, visual and vibrotactile feedbacks were provided separately and combined. Our results indicate that the users were able to assimilate the provided feedback mechanisms, with *performance* following a similar pattern across the different feedback scenarios. The association of visual and vibratory feedback was effective, with users being able to control the system with both modalities combined. The use of short pulses of vibrotactile stimulus as a spatial cue in the beginning of each trial was also effective. The angle tolerance in each scenario had a significant impact in *user performance*, highlighting the boundaries of control specificity for the system, and extreme leg positions were easier to match. Thus, this study shows that the association of VR and vibratory feedback in the back of the user are a viable mechanism to convey information in a task of spatial representation and provided insightful results to move this Biofeedback system into the clinical scenario.

Our final goal is to use this system to provide alternative feedback to lower limb amputees during preprosthetic training, bypassing the lack of sensory information regarding the lost limb and encouraging motor and functional training by recruiting the residual musculature in the stump [1].

## 2. Materials and Methods

### The system

This Biofeedback uses surface EMG to control two feedback mechanisms (Fig. 1), which are the knee joint of a prosthetic leg of a humanoid avatar in a VR environment (visual feedback) and a matrix of 16 vibrotactile actuators placed in the back of the user (vibratory feedback). The study was approved by the ethics committee at the Federal University of São Paulo (n0 3.030.944) and participants signed a consent form stating the proposal of the experiments.

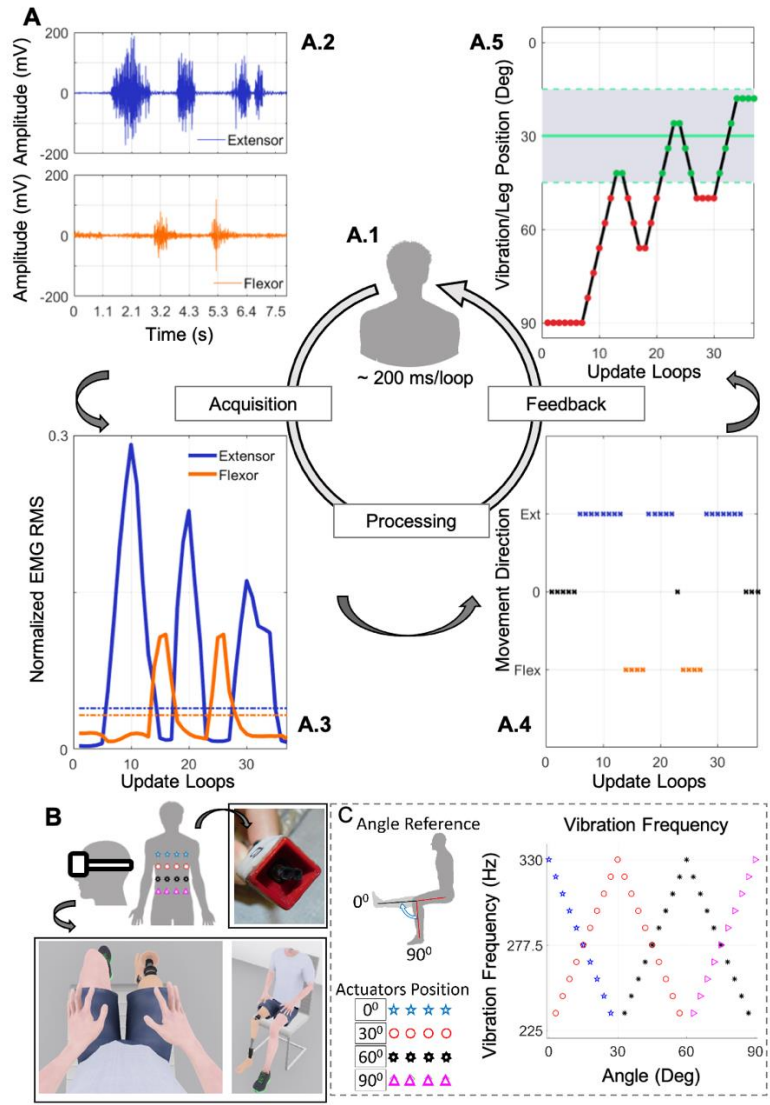

**Figure 1 – Schematic representation of Biofeedback developed in this study.** A) Overview of the closed-loop process of feedback control. *A.1*- Steps in the Biofeedback cycle; *A.2*- Surface EMG data acquisition; *A.3*- Processing of the signal; *A.4*- Signal classification and decision making according to the recorded muscle activity; *A.5*- Overall response of the system and target position (shaded region). B) Feedbacks provided to the user according to the task. View of the first- and third-person perspectives of the Virtual Reality environment, and magnified view of one of the vibrotactile actuators used in the back. C) Equivalence between the position of the virtual leg and the vibration provided in the back of the user and a graph showing the transition of vibration intensity between different rows of the vibratory matrix, using pulse-width modulation. Feedbacks are provided either separated or combined.

*Data Acquisition*

Data acquisition was inside a Faraday Cage (2.00 m x 2.00 m x 1.50 m) using an OpenEphys® acquisition board, capable of recording up to 128 channels at 30 kHz simultaneously. The board was connected to a 16-channel multiplexed differential amplifier chip from Intan technologies for the surface EMG recordings. The amplifier chip is a 16-bit analog-to-digital converter, with CMRR of 82 dB and a differential gain of 192 (Fig. 1A). The use of a Faraday cage is to improve signal quality and make the system more robust to be applied in the clinical setting, where ambient noise is harder to isolate [23].

This characterization study was performed on able-bodied participants, with no amputation, and for that the dominant leg of the user was immobilized using an orthopedic boot fixed on the chair, allowing only isometric contractions of target muscles. The surface EMG was recorded using two pairs of channels in bipolar configuration, one on the Rectus Femoris (extensor) and the other on the Semitendinosus (flexor) muscle of the participants, according to the SENIAM [24].

The interface with the acquisition board was done using a software provided by OpenEphys. The sampling frequency was set to 10kHz to allow real time control of the system, once the OpenEphys interface stores the data in packages of 2048 data points. Thus, sampling at 10 kHz allows to update the data at ~200 milliseconds, suitable for real-time applications [25].

*System Calibration*

System Calibration used the baseline and maximum voluntary contraction (MVC) levels of each person to set the minimum and maximum references of the system. The baseline was 15 a seconds recording with muscles relaxed and the MVC was three trials of maximum contraction for 6 seconds for each muscle group, followed by a 2 minutes break. We computed the RMS value for each of the three MVC trials for each muscle group (eq. 1) and averaged that value to set as the final MVC of each muscle (eq. 2).

$$_{RMS}^{m}MVC_i = \sqrt{\frac{1}{N} * \sum_{N} |u(n)|^2}$$

(1)

where, $MVC_i$ is the MVC value for that particular trial, and *m* indicates the muscle group and *RMS* indicates that this is the RMS value for that recording. The *N* indicates the total number of samples of the trial, *n* represents the current sample and *u(n)* represents the discrete voltage values obtained

$$_{RMS}^{m}MVC = (\sum_{i=1}^{3} {}_{RMS}^{m}MVC_i)/3$$

(2)

where, *MVC* represents the final MVC value, *m* indicates the muscle group and *RMS* indicates that this is the RMS value for that recording. The index *i* represents each one of the three different trials for each of the two muscle groups.

*Data Processing*

Data Processing consisted of loading and processing data packets in real-time using Matlab® 2017b, at intervals of 200 milliseconds and with an overlap of 60% to improve signal smoothness. Each data packed was filtered in the frequency range from 10 to 500 Hz and notch filtered in 60 (± 2) Hz and its harmonics. Following that, the signal was downsampled to 2 kHz and we subtracted the Root Mean Square (RMS) value of the baseline recording, to account for intrinsic baseline activity. Lastly, we use the RMS value of the data packet to drive the system.

*User Feedback*

User feedback was provided in two possible ways: the control signal was used to drive the prosthetic leg of an avatar in a VR environment, or to activate a 4-by-4 matrix of vibrotactile actuators (Fig. 1B).

In the VR environment, the user sees the avatar in first person, with a prosthetic leg matching its dominant leg. The avatar can perform the movements of knee extension and flexion with the prosthetic leg, in a range from $0^0$ to $90^0$ (Fig. 1C), controlled by Matlab via TCPIP communication protocol. There was no effect of gravity in the virtual leg.

The vibrotactile actuators are Eccentric Rotating Mass DC motors set to vibrate within a frequency range from 250 to 330 Hz, optimal for Pacinian corpuscle stimulation (40-400 Hz) [26]. The 16 motors were disposed on a 4x4 grid pattern, 5 cm apart on the back of subjects (Fig. 1C) and held in place using thin and adjustable tissue stripes. Motors were activated using an Arduino Mega board, and vibration frequency was controlled via pulse-width modulation (PWM). Each row had a $30^0$ overlap with its adjacent, to produce a continuity effect during transitions, and peaking in intensity at $0^0$, $30^0$, $60^0$ and $90^0$ (Fig. 1C). Outside of these peak positions, the PWM allows a smooth transition, with gradual activation of the adjacent rows. These peaking positions were also used as the vibratory cues for the participants during the proposed tasks.

The choice of having 16 motors aligned in a 4x4 grid was based on preliminary testing, which showed that, because of the reduced spatial acuity and the disposition of the dermatomes in the back, [27] users differed in terms of sensitivity in different areas. This arrangement allowed us to avoid this problem, stimulating all the motors across each row.

*Driving the System*

To identify a movement intention, two criteria were stablished. First, a threshold was initially set at two times the RMS value of the baseline in rest for each channel (these thresholds are later changed during training). Only values above the threshold were considered as a movement intention. Second, to trigger a movement, the activation level of the antagonist muscle must be smaller than 80% of the activation of the agonist. This condition requires the user to relax the antagonist muscle in order to trigger a response in the system. The resulting control paradigm is that the feedback responses would only change when there is an intention to move and remain in the same position if no muscle groups are recruited. The choice of 80% difference in amplitude between signals was based on our experience during pilot tests. This is another parameter that can be later used as a method to train isolated muscle recruitment during clinical use.

When both criteria to detect a movement are met, the virtual prosthetic leg will respond moving at a fixed speed, determined by the percentage value of the agonist muscle RMS when the threshold was crossed. To reset the virtual leg speed, the user must reduce its muscular activation below the threshold and cross it again, by relaxing the muscles. The available leg speed values in respect to the RMS percent were 8, 10, 15, 20 and 30 degrees per update loop, and the percentage RMS range for each of the speeds is equally divided within the region above the threshold value and the maximum activation. This maximum activation is determined in the calibration step, when we calculate the user MVC. This way, the muscle contraction is represented as a percentage of the total activation, eliminating any possible imbalance between the absolute values of activation from the Extensor and the Flexor muscles, which could result in one of the movements being particularly faster than the other. For the vibrotactile feedback the idea is the same, but instead of the leg angle, what changes is the position where vibration is applied in the back.

The choice of a fixed speed is because having a proportional speed according to the RMS value made the device much harder to control, because people present different MVC levels. With that, we assure that leg speed would be proportional to the activation level during the onset of the contraction, and users can sustain a lower level as long as it surpasses the established threshold. This strategy is tailored to balance the use of the system to people who might present difficulty to sustain muscle contraction at higher levels for longer periods of time, which is what we expect in the amputee subjects, and we implemented that in order to give them the best sense of agency possible.

**Biofeedback Experimental Protocol**

*User Training*

User Training was the last step of system calibration, setting the thresholds for each muscle using a visual training interface. It consisted of moving bars representing each of the surface EMG channels (Fig. 2A). Following each of the three trials, users reported the amount of effort associated with each movement, using the Borg CR10 scale [28] and if there was any difference in effort between both movements. From that, the threshold values were rebalanced to reflect an equal effort for both movements (knee extension and flexion).

In the second part of the training, using the VR, the users performed two similar tasks. First, only moving the virtual prosthetic leg, and later, moving the virtual prosthetic leg along with the vibratory feedback (Fig. 2A).

*Tests*

*Testing* was performed on 13 able-bodied subjects (m: 6, f: 7), with no amputation, with age $24.7 \pm 2.5$ years and mean Body Mass Index of $24.5 \pm 4.2$. *Inclusion criteria*: age between 20 and 30 years old. *Exclusion criteria*: history of traumatic lesion in the lower limbs and motion sickness symptoms in VR.

During the experimental protocol, each person performed a total of 9 tasks (Fig. 2C). Each task had 8 trials, composed of: a cue beep, a silent period (1s), a vibratory stimuli (1s), another silent period (2s), another beep and the beginning of the trial (up to 20s) (Fig. 2D). Each stimulus was a random activation of a row of motors. For each task, there was a total of two trials in each position, corresponding to $0^0$, $30^0$, $60^0$ and $90^0$ of knee joint (Fig. 2B). The goal was to discriminate the received stimuli and respond accordingly in each scenario, controlling the system with the provided feedback.

These feedbacks were: *VR* (visualization of the virtual leg movement only); *VIB* (vibrotactile stimuli only, no visual feedback); and *VR+VIB* (visualization of the virtual leg movement with an associated vibrotactile stimuli) (Fig. 2B). In scenario *VR*, after each vibratory stimulation, the user had to position the virtual leg where the vibration indicated. In scenario *VIB*, the goal was to place the vibratory feedback in the region corresponding to the vibratory stimuli, without visual aid. Lastly, in scenario *VR+VIB*, the user received the vibratory cue and used both feedback modalities simultaneously to perform the task. The feedback scenarios were sorted random and uniformly within users. The sequence of cues (vibratory stimuli) was also random. After performing one task with each feedback, the difficulty was increased without the user knowing. This was achieved reducing the tolerance for movement representation, and the difficulty levels were *Difficulty 1*, *Difficulty 2*, and *Difficulty 3*, which corresponded to a deviance of $\pm 15^0$, $\pm 10^0$, and $\pm 5^0$ from the target angle, respectively (Fig. 2B).

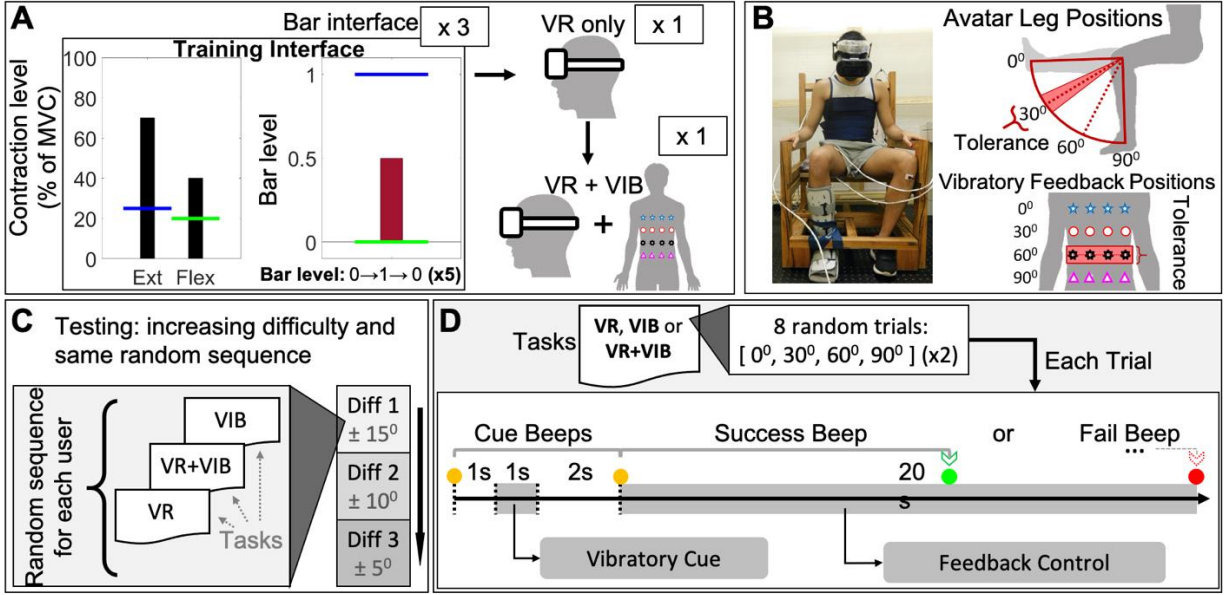

**Figure 2 – Schematic representation of the experimental protocol.** A) Sequence of feedbacks for each step in the training session. Users start interacting with a graphical interface, to understand the mechanisms to control the system, while threshold values are calibrated. Following, they are trained to move a virtual leg inside the virtual reality environment, and lastly, vibratory feedback is added to the leg movements in the final training. B) Description of the feedback modalities and picture of a subject performing the tasks. Subjects performed a total of 9 tasks with 8 trials each and responded using the feedback modality provided in that task (VR, VIB or VR+VIB). C) Sequence of the tasks during the protocol. The random sequence of feedbacks is maintained as difficulty (Diff) is increased. D) Description of the steps involved during each trial within a task.

## User performance

We defined *user performance* as the ratio of correct trials over the total (eq. 3).

$$User\ Performance = \frac{correct\ trials}{total\ trials} \qquad (3)$$

where, total trials = 8, and correct trials is defined as (eq. 4)

$$correct\ trials = \sum matched\ positions \qquad (4)$$

*User performance* is a binomial variable, which can take the values of 0, 0.125, 0.25, 0.375, 0.5, 0.625, 0.75, 0.875 and 1.0.

## Time score

We proposed the calculation of a *time score* (eq. 5), given by

$$time\ score = 1 - \left(\sum_i^n t_i^{trial}/T\right) \qquad (5)$$

where $i$ is the trial index, $n$ is the number of trials, $t_i^{trial}$ is the time spent on each trial (maximum 20 seconds) and $T$ is the total time of the task (160 seconds). This score reflects user *performance* in terms of the time required to complete each task.

## Statistics

*User performance* and *time score* values were tested via Kruskal-Wallis test followed by Dunn-Sidàk post-hoc test. Here, *user performance* is defined as the ratio of correct trials over the total (8 maximum).

*User performance* for *Inner* angles considers $30^0$ and $60^0$, and for *Outer* angles $0^0$ and $90^0$. Data was compared according to the difficulty level (*Difficulty 1*, *Difficulty 2* and *Difficulty 3*) for the same feedback modality, and by feedback (*VR*, *VIB* and *VR+VIB*) for the same difficulty. Mann-Whitney test was used to compare the *performance* between *Inner* and *Outer* angles for the same combination of feedback and difficulty. Absolute values (solid bars, named as "uncorrected proportion of users") were corrected to estimate confidence intervals [29], used to compare the proportion data (colored bars with confidence intervals) in Fig. 3C and Fig. 4C. All tests used a significance coefficient of α=5%.

## 3. Results

**User Performance**

Analysis of *User Performance* shows no statistical difference in average *performance* between different feedback types for a given difficulty level (Fig. 3A). Comparing amongst the difficulty levels for each feedback modality, *performance* dropped significantly for *Difficulty 3* in all cases (Fig. 3A).

We grouped the data in trials of *Inner* ($30^0$ and $60^0$) and *Outer* ($0^0$ and $90^0$) angles to assess whether there was any difference in *performance* for these positions (Fig. 3B). Performance was significantly smaller for trials of Inner angles (Fig. 3B). No statistical difference between tasks was found within trials of *Outer* angles (Fig. 3B.1), and trials of *Inner* angles presented a similar statistical pattern of *performance* when compared with all possible angles (Fig. 3B.2).

We conducted an analysis of user success (Fig. 3C), which compared the percentage of users that scored above 75% in the trials. To do so, users must correctly identify a minimum of 6 out of 8 positions, including at least half of the *Inner* angles, which were significantly harder to match than the *Outer* angles (Fig 3B). Once again, *performance* dropped significantly in all scenarios as users reached the *Difficulty 3*.

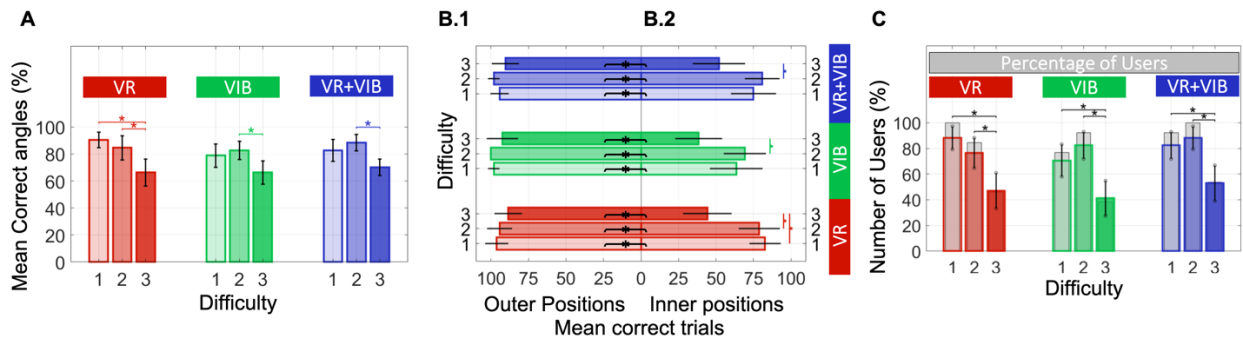

**Figure 3** – *User performance* analysis. A) Performance testing considering each difficulty level grouped by feedback type. B) Performance testing considering trials of *inner* ($30^0$ and $60^0$) and *outer* ($0^0$ and $90^0$) angle pairs and grouped by feedback types. C) Percentage of users who presented a Performance ≥ 75% and confidence intervals. Solid bars (grey) represent absolute (uncorrected) proportions of users.

**Scoring Analysis**

We used the *time score* (1) index to test whether the feedbacks have an influence on the time required to complete each task and how users would perform in the scenario with both feedbacks in respect to the others with a single feedback.

The idea behind the *time score* metric is to add another layer of information, in this case the "time to complete the task" (1 task = 8 trials). This metric is useful to highlight the scenario where the user completed the trials faster on average, which results in a higher *time score* value. Setting a threshold allowed us to quantitatively compare how users performed in the tasks.

In that sense, if someone succeeded in *performance*, matching 6 out of 8 angles, the *time score* also considers how long the user takes to do so. We set the threshold value at 0.5 because it corresponds to 10 seconds, which is half of the total time available to complete each trial within a task. It is also double of the median time users needed to complete each trial within the tasks, which was close to 5 seconds.

In respect to the statistical testing of the *time score*, we decided to keep this metric because it allowed us to add more information to analyze the effects of the feedbacks and the task difficulty, as is shown in Fig. 4 and Fig. 5.

The analysis of *time score* shows no statistical difference comparing different feedbacks for the same difficulty (Fig. 4A). Statistical difference is observed between *VR* and *VR+VIB* feedbacks, comparing the *Difficulty 3* level against *Difficulty 1* and *Difficulty 2* (Fig. 4A), now showing that users not only had a decrease in *performance*, but also took longer to complete the tasks in *Difficulty 3*.

As difficulty increased, the *time score* of users varied largely and followed no specific pattern (Fig. 4B). An exception is the *VR+VIB* feedback in *Difficulty 2*, where all users scored above 0.5, which means that they completed the task in less than 80 s (Fig. 4B). We used that condition as a threshold and compared the percentage of users that matched this criterion in the other conditions as well (Fig. 4C). We observed that, for *Difficulty 2*, significantly more users met the criterion in the task using *VR+VIB* feedback when compared to *VR* or *VIB* feedbacks alone (Fig. 4C). This result suggests that, although there was no significant difference in *performance* between the three scenarios for *Difficulty 2*, the combination of both feedbacks allowed users to complete the tasks faster.

Finally, Fig. 5 presents scatter plots of *user performances* against *time scores*, with the regions corresponding to the thresholds depicted as the shaded area. It summarizes the information presented in Fig. 3C and Fig. 4C and helps to visualize intrinsic characteristics associated with each feedback modality in respect to the differences amongst users.

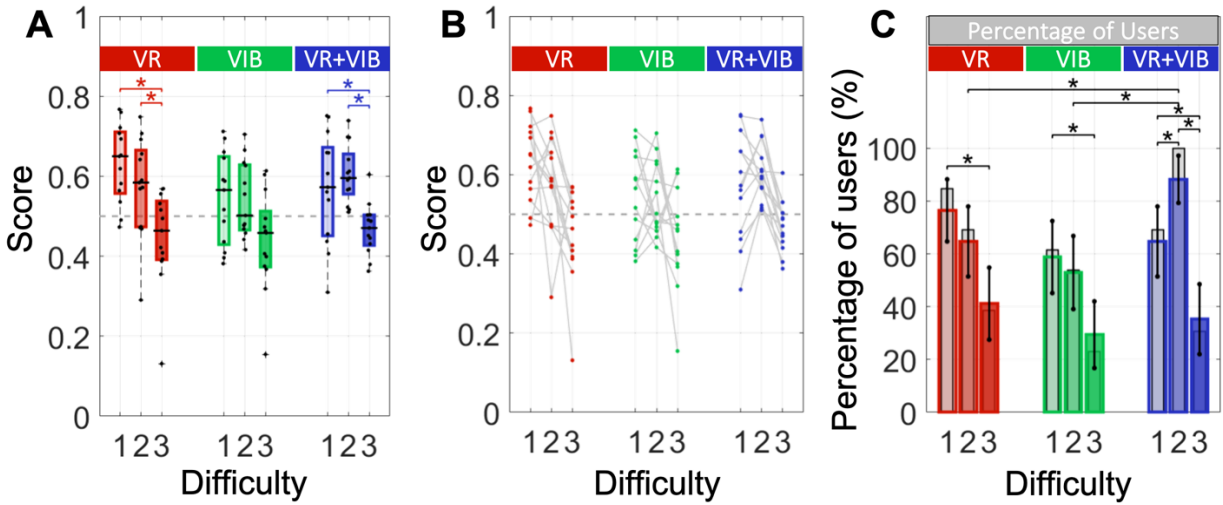

**Figure 4 – Analysis of *time score*.** A) Median comparison of *time scores* among users, grouped by feedback types. B) Representation of *time score* differences for each user in each difficulty level, grouped by feedback types. C) Percentage of users who presented a times core ≥ 0.5 and confidence intervals. Solid bars (grey) represent uncorrected proportions of users.

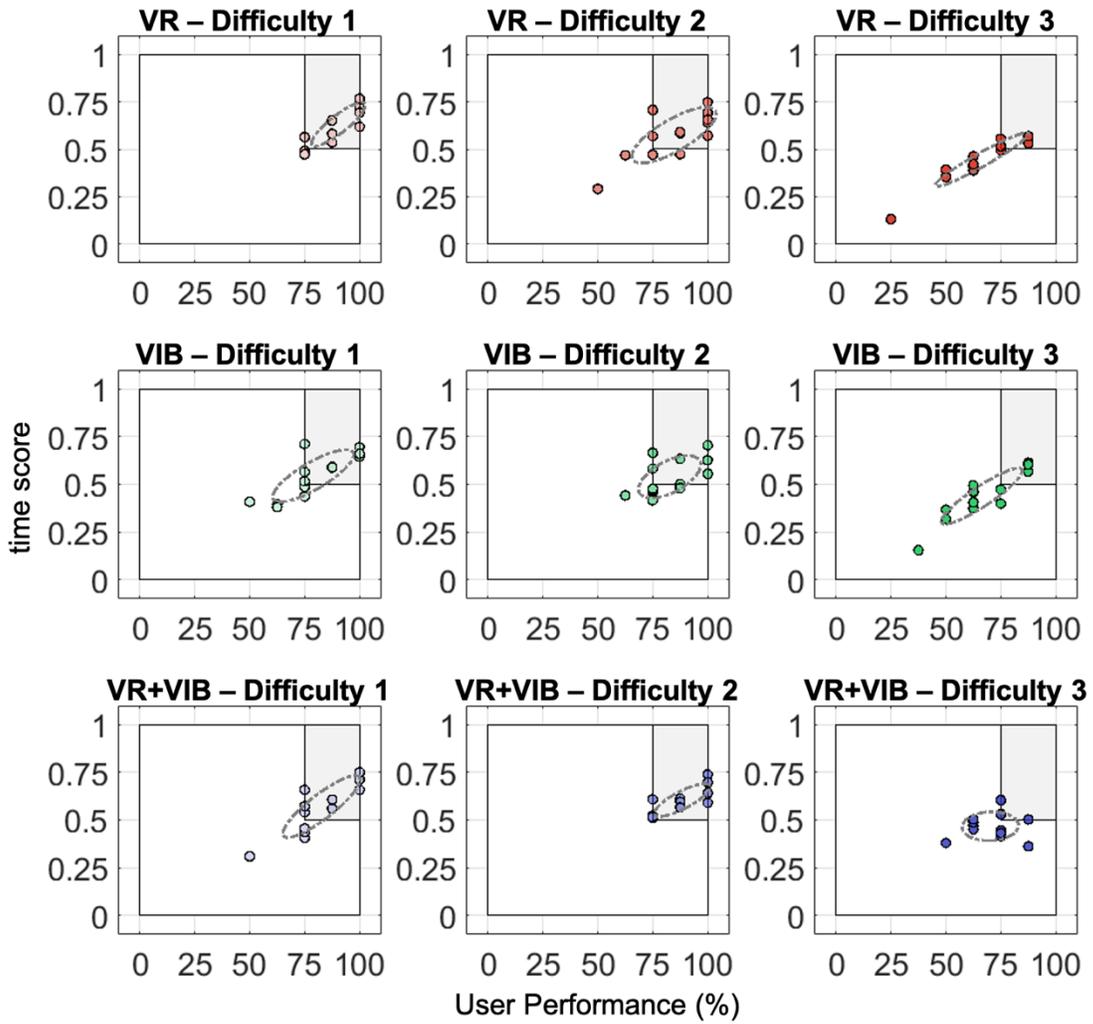

**Figure 5 – Scatter plot of Performance versus *time score*.** Each plot represents a combination of feedback type and difficulty, and each dot corresponds to a different user. Shaded areas represent the values of Performance ≥ 75% and *time score* ≥ 0.5.

## 4. Discussion

In this study we proposed a new Biofeedback device intended to be used on the Preprosthetic rehabilitation phase for people with transfemoral amputation. With this new system, we propose a mechanism to bypass the lack of sensory information from a virtual prosthesis and help them to assimilate visual and vibrotactile stimuli as a cue for movement representation [13]. Here we try to answer two questions: (*i*) Is the user capable to assimilate simultaneously visual and vibrotactile information to represent the movement of a prosthetic leg? (*ii*) Does the association of both feedbacks affect the *performance* of the users within a single session protocol?

To answer that, we conducted a series of tests with able-bodied participants, simulating the similar expected conditions for transfemoral amputees [1,4]. We immobilized the dominant leg of the user, allowing only isometric contractions of the target muscles, used to drive both available feedback modalities: a prosthetic leg designed in a virtual reality environment and a 4-by-4 matrix of vibrotactile actuators. During the protocol we varied the type of provided feedbacks (visual only, vibrotactile only, or both) and progressively increased the difficulty of each task.

Our results show that all users were capable to recognize both feedback modalities, both separate and combined, being able to respond accordingly throughout the tasks. We also show that, for a one-session protocol, the last difficulty level imposed greater challenge for most users, explained by the significant drop in *performance* disregarding the feedback modality itself.

**Task Difficulty**

The increase of task difficulty is evidenced in both aspects: *user performance* and *time score* (Fig. 3, Fig. 4). It showcases that, although users already had experienced the protocol, acquired on earlier difficulties (*Difficulty 1* and *Difficulty 2*), only knowing what to expect from the tasks was not enough to keep their *performances* at a higher standing. Complimentarily, their *performance* index did not differ amongst the feedback types for the same difficulty. It supports the conclusion that *Difficulty 3* was equally more challenging in all feedback modalities.

Furthermore, we see that trials of inner angles ($30^0$ and $60^0$) were significantly more difficult, contributing to the observed differences in *performances* (Fig. 3B). These two intermediate positions contained most of the information regarding task *performance*, since for outer angles the leg would stop moving at the extremities ($0^0$ and $90^0$), making it easier to achieve. Thus, when users had to reproduce such positions, they would just extend or flex the leg until it reached these extremes, and the angle tolerance became less relevant. In contrast, when trials consisted of inner angles, this strategy would not work, and users were faced with the challenge of matching the proposed angle positions.

Additionally, the *time score* of users, depicted in Fig. 4B vary largely in respect to the different conditions. Is worth mentioning that the notably smaller values for each combination of feedback and difficulties correspond to different users, who performed poorly in those specific scenarios. Whereas the *time score* of some users increased from *Difficulty 1* do *Difficulty 2*, others presented a substantial decline. This shows that the apparent outliers in Figure 4 A,B was not a single user performing poorly, but instead different users who performed poorly in different conditions. It must has occurred either because the increase in difficulty was such that users could not match the target positions with the provided feedback, or because users did not expect a change in task difficulty. Although the second argument is very plausible, this decline in *performance* did not occur for *VR+VIB* feedback. All users progressed, or had a small decline using the *VR+VIB* feedback, pointing towards its greater efficacy against the isolated feedbacks *VR* and *VIB* only.

**Scoring Analysis**

From our knowledge, this protocol is the first controlled study using this type of Biofeedback system, providing insights on how users would perform using different feedback modalities under different level of task difficulties. The main idea was to guide new protocols on how the feedbacks would be perceived by the users, and to what extend success in the tasks was achievable during different task difficulties. To accomplish this, we had to bypass few aspects regarding the operant conditioning strategies [16]. In especial, because protocols must obey the learning curve of each user, progressing in difficulty only when users match the established criteria (e.g. minimum *performance*, or reach a certain threshold) [16]. This is also true in the case of Motor Learning, which is a process that involves several factors, such as the difficulty, motivation, feedback type and practice over time [30].

We reasoned that this would be particularly difficult to achieve in a single session, and to compare [16]. And it is important to mention that users had no previous contact with the proposed tasks, and they were not aware that the difficulty was gradually increased throughout the trials.

Therefore, accounting that users would present different learning curves for the proposed tasks, we carried an analysis on the percentage of users who scored correctly 75% or more along the trials for each task. This criterion requires that the user matches at least two of the four intermediate presented angles, shown to be more difficult to reproduce. The idea was to identify how many of the users could perform successfully for each task, and if there was any condition that would prevail in respect to the others.

The observed result reflects the same aspect observed in the analysis of *performance* and *time score*, which showed that *Difficulty 3* was more difficult than *Difficulty 1* and *Difficulty 2*, with significantly less users scoring on 75% or above (Fig. 3C). Additionally, we see that all users achieved the established criteria in *Difficulty 1* with the *VR* feedback, and in *Difficulty 2* with the *VR+VIB* feedback (Fig. 3C). Despite, the confidence intervals do not allow us to infer that these conditions were significantly better for all users, it is very suggestive that the visual modality, given an easier scenario (*Difficulty 1*), is more informative than the others (*VIB* and *VR+VIB*). Also, as the difficulty increased (*Difficulty 1* to *Difficulty 2*), all users performed above 75% for the *VR+VIB* feedback, and not anymore in the *VR* only.

Following the same approach, we conducted an analysis of minimum score, since all users completed the task in less than half of the time (*time score* ≥ 0.5) in *Difficulty 2* with the *VR+VIB* feedback. The idea was to compare how this condition differed from the others and evaluate if the influence of time would be a determinant factor to discriminate all scenarios more effectively. We observed that, for all feedback types, a significantly smaller number of users met this criterion in *Difficulty 3* (Fig. 4C), which is also consistent with the results observed so far.

Significantly more users met the criterion with the *VR+VIB* feedback in *Difficulty 2*, when compared with the *VR* or *VIB* isolated (Fig. 4C). The same does not happen in the case of *VR* feedback in *Difficulty 1*, where all users also had *performance* ≥ 75%, but some users failed to achieve a *time score* ≥ 0.5. We can see a summary of these results in Fig. 5, with a representation of the *performance* versus the *time score*, showing how the results scatter for each condition. The highlighted region represents the zone that meets the criteria established in both thresholds (*performance* ≥ 75% and *time score* ≥ 0.5), and the ellipse represents the covariance of both variables. We notice that all users fall within that region for the *VR+VIB* feedback in *Difficulty 2*, and how *performance* and *time score* are intrinsically correlated, but the later entail some aspects regarding the tasks that the first does not.

In summary, the visual feedback was enough for all users to perform satisfactorily in the easier scenario (*Difficulty 1*), but users differed in the time required to complete the task. Also, the presence of both visual and vibrotactile feedback did not result in any significant improvement at that moment (*Difficulty 1*). However, as the tasks progressed, and difficulty increased, users started to assimilate and rely on the vibrotactile feedback, using this information to perform better and faster during the proposed tasks.

Cuppone et al. also found a similar result in a visuo-spatial task with added vibrotactile feedback, showing that early decreased *performance* in users with added feedback is related to the sensorimotor integration of the vibratory information in the process of spatial localization [13]. In other words, is likely that users need some time to assimilate the auxiliary feedback modality, and after a few iterations, *performance* is better using congruent feedback than control groups [13].

We believe that longer training periods are necessary in order to produce an effect on *performance* in *Difficulty 3*. Also, we must consider the learning curve of users, increasing the difficulty in respect with individual progress. Nevertheless, the tests highlighted the capacity of users to assimilate the information in *VR+VIB* feedback, emphasizing the possibility of using vibrotactile information as a supporting mechanism for spatial localization in the proposed tasks.

Complimentarily, the *VIB* feedback alone did not prevail in any of the aspects considered, even though it was the same modality used to provide the cues for spatial localization during the tasks. These results are congruent with the literature describing the predominance of visual feedbacks for spatial localization tasks [31], and that vibratory feedback can be incorporated as an accessory feedback modality [8,9].

**Integrated Biofeedback in Rehabilitation**

It is important to emphasize that the proposed Biofeedback system is designed to be used as a rehabilitation tool for people with transfemoral amputation, but first we must attest that the system is ready to be applied in that population. Therefore, our objective here is to showcase that the system, the protocol and feedback mechanisms are secure, efficient and reliable. The full extent of the effects of the Biofeedback training can only be assessed after the evaluation with the amputees, and the results obtained so far stablish a safe and operational protocol to accomplish it.

For these people, the experimental protocol will be provided as a complimentary therapy and adapted from the proposed tasks, in order to follow a longer and progressive training period. In this sense, it is expected that with a continued use the observed effects will be more evident, as literature supports [13]. A recent study showed the benefits of long-term training using Brain-Computer interfaces (also referred to as Neurofeedback), and even in a scenario such as spinal cord injury, patients were able to assimilate virtual legs as their own [11]. Additionally, promising effects were also obtained in a series of other conditions [12,18,19], showcasing the potential benefits associated with these systems in the clinical rehabilitation scenario. We also believe that motor learning will have a significant contribution in the learning process, once users start to familiarize with the aspects involved in muscle recruitment and the proposed tasks [16].

The goal of combining the crossmodal sensory feedbacks with the association of *VR+VIB* stimulus was to provide a better framework to perform the proposed tasks, and other authors have applied similar feedback mechanisms in other studies, such as vibratory information to represent grasping force and aperture in myoelectric prosthesis [9]. Furthermore, crossmodal feedback strategies take advantage of the multisensory nature of the human brain [6], conveying information to the subject via an idle sensory modality - in this case, the mechanoreceptors in the surface of the skin in the back. This strategy was also applied in other scenarios, such as association of visual and auditory stimuli, [10] as well as crossmodal activation of the visual cortex during Braille reading. [8]

For future studies we intend to study the association of visual and tactile information in the cortex and evaluate if extensive training can result in a representation of the vibratory information on the somatosensory leg area. For that, we designed this setup to be readily electroencephalography-compatible, with which we intend to evaluate changes of specific rhythms in the motor cortex such as the sensorimotor rhythm, commonly associated with motor planning and execution [16]. We also speculate that, considering the anatomical mapping of the somatosensory cortex, the proximity of the regions associated with the back and the leg could have a positive influence in the flow of information to the areas that correspond to the leg. If this is the case, then the association of visual and vibratory feedback is indeed a valuable tool to be used during preprosthetic rehabilitation, targeting brain areas undergoing plasticity after amputation, and potentially reducing the effects of phantom limb phenomenon. Lastly, we also expect to help provide a better mechanism for the embodiment of prosthetic devices, fulfilling the lack of sensory information for the users, which is one of the main reasons for prosthesis abandonment [5].

**5. Acknowledgements**

This work is supported by the CNPq proc. n$^0$ 442563-2016/7, the INNT and CAPES Proex program (MCTI, Brazil). The authors thank the volunteers, especially André Posso and Fabiano Vittoretti, and Professors Henrique Amorim, Adenauer Casali, Karina Casali and Matheus Moraes for supporting the development of this study.

## 6. Author Biography

**João Vitor da Silva Moreira**

M.Sc. in Neuroscience (Fed. Univ. of São Paulo, Brazil) and B.Sc. in Biomedical Engineering (Vale do Paraíba Univ., Brazil). PhD candidate in Neuroscience (SUNY Downstate Medical Center, USA).

**Karina Aparecida Rodrigues**

MSc in Mechanical Engineering (São Paulo State Univ., Brazil), BSc. in Physiotherapy (Taubaté Univ., Brazil) and PhD candidate in Neuroscience (Federal Univ. of São Paulo, Brazil).

**Daniel José L.L. Pinheiro**

MSc. in Neuroscience and BS in Biomedical Engineering (Federal Univ. of São Paulo, Brazil). PhD candidate in Neuroscience (Federal Univ. of São Paulo, Brazil)

**Thaís Cardoso Santos**

Biomedical Engineer and BSc in Science and Technology (Federal Univ. of São Paulo, Brazil). Electronics Technician (Technical School Everardo Passos, Brazil).

**João Luiz Vieira**

B.Sc. in Science and Technology (Federal University of São Paulo, Brazil). Currently working as a Game Developer (Softgames Berlin, Germany).

**Esper A. Cavalheiro**

Esper A Cavalheiro, MD, PhD, Senior Professor in the Department of Neurology and Neurosurgery (Escola Paulista de Medicina, Federal Univ. of São Paulo, Brazil)

**Jean Faber**

Adjunct Professor (Federal Univ. of São Paulo, Brazil). PhD & MSc. in Quantum Information (National Lab. for Scientific Computing-MCTI, Brazil). BSc. in Physics (Fed. Univ. of Juiz de Fora, Brazil).